# A Novel Multidimensional Reference Model For Heterogeneous Textual Datasets Using Context, Semantic And Syntactic Clues

Ganesh Kumar
Computer and Information Sciences Department (CISD),
Universiti Teknologi PETRONAS, Bdr Sri Iskandar, 32610
Seri Iskandar, Perak, Malaysia
Ganesh_17005106@utp.edu.my

Shuib Basri
Computer and Information Sciences Department (CISD),
Universiti Teknologi PETRONAS, Bdr Sri Iskandar, 32610
Seri Iskandar, Perak, Malaysia
Shuib_basri@utp.edu.my

Abdullahi Abubakar Imam
School of Digital Sciences, Universiti Brunei Darussalam,
BE1410, Brunei Darussalam
abdullahi.imam@ubd.edu.bn

Abdullateef Oluwagbemiga Balogun
Computer and Information Sciences Department (CISD),
Universiti Teknologi PETRONAS, Bdr Sri Iskandar, 32610
Seri Iskandar, Perak, Malaysia
abdullateef.ob@utp.edu.my

Hussaini Mamman
Computer and Information Sciences Department (CISD),
Universiti Teknologi PETRONAS, Bdr Sri Iskandar, 32610
Seri Iskandar, Perak, Malaysia
hussaini_21000736@utp.edu.my

Luiz Fernando Capretz
Department of Electrical and Computer Engineering
Western University, London, Canada.
lcapretz@uwo.ca

*Abstract*— With the advent of technology and use of latest devices, they produces voluminous data. Out of it, 80% of the data are unstructured and remaining 20% are structured and semi-structured. The produced data are in heterogeneous format and without following any standards. Among heterogeneous (structured, semi-structured and unstructured) data, textual data are nowadays used by industries for prediction and visualization of future challenges. Extracting useful information from it is really challenging for stakeholders due to lexical and semantic matching. Few studies have been solving this issue by using ontologies and semantic tools, but the main limitations of proposed work were the less coverage of multidimensional terms. To solve this problem, this study aims to produce a novel multidimensional reference model using linguistics categories for heterogeneous textual datasets. The categories such context, semantic and syntactic clues are focused along with their score. The main contribution of MRM is that it checks each tokens with each term based on indexing of linguistic categories such as synonym, antonym, formal, lexical word order and co-occurrence. The experiments show that the percentage of MRM is better than the state-of-the-art single dimension reference model in terms of more coverage, linguistics categories and heterogeneous datasets.

## I. INTRODUCTION

"Big Data" refers to data sets with sizes beyond the ability of commonly used software tools to capture, curate, manage, and process data within a tolerable elapsed time. Various industries with heterogeneous data are facing problems related to storing, managing, retrieving and analyzing of large amount of data. Big Data plays an important role in retrieving useful information from the large datasets with the help of advanced tools and algorithms[1]. Nowadays, data produced in formats such as structured, semi-structured and unstructured data from a multidimensional nature of resources and applications that cannot be processed through simple tools [2].

In general, Big Data can be explained according to three V's: Volume , Velocity and Variety [3]. Also, the other characteristics of Big Data described in [4] are volume, variety, velocity, veracity, valence, and value. Later on, in [5] 10V's volume, variety, velocity, veracity, variability, viscosity, volatility, viability, validity, and value are exposed.

In Big Data Variety, the heterogeneous types of data formed, and it further classified in three types namely, Structured, Semistructured and Unstructured (SSU) [6],[7]. Structured data is organized data in a predefined format and stored in tabular form whereas semi-structured data is a form of data which cannot be queried as it does not have a proper structure which confers to any data model and unstructured data is heterogeneous and variable in nature such as text, audio, video, and images. Due to heterogeneous data, it cannot be processed with simple tools and techniques which creates the problem heterogeneity and similarity matching [2] in result, decision maker cannot make decision based on scattered data.

With the advent of the technology, the computers are nowadays used to retrieve the linguistics information from textual data which is known as Computational Linguistics (CL)[8],[9]. CL is classified into many categories but among them context clues[9], semantic[10], and syntactic [11] matching is widely used in the domain of linguistics. CL helps in identifying and matching [12] of related words from input datasets with the data dictionary (domain knowledge).

The domain knowledge further known as reference model (RM) have been used in the field of NLP and semantic-lexical matching. Vasilieous et al. in [13]–[15], proposed a single dimensional reference model (SRM) for medical data quality of textual dataset. The SRM only matches one token to one term at time and it was developed for structured dataset whereas the same patient's data can be represented in other forms of terms. Also, in other formats (semi-structured or unstructured). Therefore, this paper proposes a multidimensional reference model (MRM) for one token to many terms matching and as well as for heterogeneous datasets.

The concept of multidimensional reference was adopted from [16],[17], in which different schemas for one to one and one to many queries for NoSQL Injection were proposed.

For further understanding about multidimensional reference model for heterogeneous textual datasets this paper is organized as follows: Section 2 and 3 describe the related work and methodology adopted for creating the MRM and experiments conducted on heterogeneous datasets, section 4 discusses the results for heterogeneous datasets and in the last section, the conclusion and future work are briefly described.

## II. RELATED WORK

Ordinarily, the reference model works as a procedure that contains the domain knowledge and relevant indexing of a topic or information of interest. It works as a common template for structured data that contains a set of parameters which are important for generating the domain knowledge [14]. The proposed multidimensional reference model as shown in Figure 8. It comes with extra features to handle heterogenous datasets. It uses a generalized natural language concept and domain knowledge which helps the input datasets in selection of appropriate multidimensional domain data. Multidimensional indexing is also an added technique which classifies linguistic words into context, semantics, and syntactic clues.

These three categories aim to assist in building the vocabulary and understanding the domain knowledge with respect to meaning, structure and representation of words as opposed to the existing reference models where the selection of terms is solely based on one-to-one relationship (please see Fig. 1).

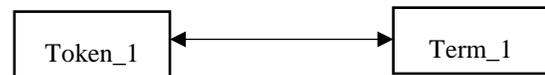

Fig. 1. Reference Model Method

It can be observed from Fig. 1 that the relationship is one-to-one. Meaning, for any given token, a corresponding match term is retrieved from the reference model. This selection is based on threshold values to identify the best matching term in the corpus. The term with the highest value is selected as a candidate for data curation.

On the contrary, the proposed multidimensional reference model utilizes a multifaceted concept as depicted in Fig. 2 below. Basically, the MRM checks the relationship between the token and its related term in multiple dimensions in order to identify the most appropriate term. Please refer to Fig. 2 and Fig 8 (Appendix A)

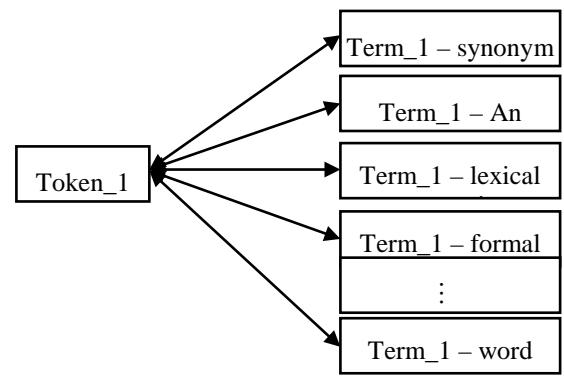

Fig. 2. Proposed Multidimensional Reference Model Core Concept

As illustrated in Fig. 8 (Appendix A), we can see that a token from the input dataset is matched with its potential related term in several dimensions such as synonym [18], semantic, lexical [19], etc. For instance, a token "bank" could score high when matched with a term "boundary", which means the edge of a river. However, if the lexical matching of same word is conducted, a financial institution, or storage may be flagged off. Therefore, it's very important to view one token in different dimensions. This will significantly increase the accuracy of terms matching at different levels of data harmonization.

The categories mentioned in MRM are context clue, semantic and syntactic. Context clues are further classified into synonym and antonym. Sample words and their score are presented in Appendix C and D. The second and most important category used in MRM for indexing the linguistics words is semantics. It plays a vital and significant role in understanding the information related to datasets.

As mentioned earlier, the first type of semantic clue is formal semantics which uses techniques such as logic, philosophy, and math to analyze data within the relationship of language and reality, truth, and possibility. The list of words and their score can be found in Appendix E and F.

The third and last category of MRM is syntactic clue which focuses on the word order and co-occurrences. In order to identify patterns amongst data points (words), the order and co-occurrences are adopted and implemented. The list of words for both the order and co-occurrences is offered in Appendix G and H.

It's important to highlight here clearly that the categories of MRM such as contextual, semantic, and syntactic clues and their score (as shown in Appendix C-H) helped in developing the multidimensional (indexed based) reference model. The MRM provides the input to the section that performs the data harmonization process. The section contains terminology extraction, rules definition, lexical matching and semantic matching which are responsible for producing data harmonization report and harmonized dataset.

## III. METHODOLOGY

For development of multidimensional reference model, following four steps have been taken (i) defining the generated tokens (ii) identifying the root word (iii) Determining the dimensions (iv) aggregating the dimensions root word. These steps are also shown in the Fig. 3.

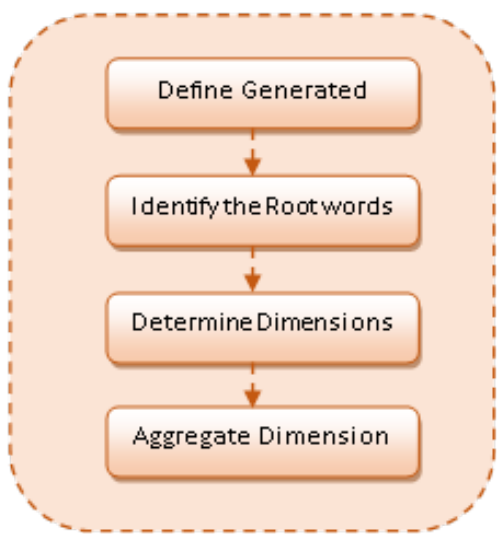

Fig. 3. Steps carried out for MRM Development

In step one, the tokens are generated from heterogeneous datasets i.e., ACE2020, Aquaint and Sarcasm news, UoA and HUA. The input datasets contain news of the daily life including sarcasm (keys and values). Participating datasets are in structured (Xls), semi-structured (JSON) and unstructured (Txt). After preprocessing the input datasets, structured dataset is formed which have been used for token generation.

In second step the root words are identified based on the generated tokens. In the third and fourth steps the determining the dimensions and aggregating them into categories of root words are formed. As stated above, the indexing scheme of dimension follows the concept of one-to-one and one-to-many cardinalities from SQL.

## IV. RESULTS AND DISCUSSION

The proposed Multidimensional Reference Model (MRM) was developed using linguistic word categories i.e., context, semantic and syntactic clues. The main aim of developing MRM is to improve the quality of terms-matching by referring to the target terms in different dimension. This is achieved with the help of indexed based domain knowledge to root-words/tokens. Indexing is generated and classified using synonyms, antonyms, lexical semantics, formal semantics, word-order, and co-occurrence.

Each word has its respective score (mrm_score()) that is empirically assigned which helps in matching terms based on defined rules, semantic, and lexical matching. The total number of words generated from linguistic word categories (i.e., context, semantic and syntactic clues) for MRM repository is 37321.

The performance of proposed MRM with existing single dimensional reference model (SRM) is compared and presented in this section. Five different heterogeneous datasets namely, ACE 2020, Aquaint, Sarcasm, HUA, and UoA are implemented on both SRM and MRM in order to obtain a justifiable conclusion on which reference model is actually better. It's important to mention here that SRM was implemented in a similar comparison on two out of the five aforementioned datasets (i.e., HUA and UoA). This indicates that our comparison is more rigorous in nature as it covers all data structures (heterogenous, to be precise).

The experiment was conducted five times (Batch 1-5) for each dataset. Batch 1 utilizes 20% of each dataset, and continuously increases 20% for the subsequent batches until 100% of each dataset is tested. This is done for both MRM and SRM to evaluate their individual performances. The batches and their respective data distributions are explained in Error! Reference source not found. The Table.1 (Appendix B) shows the results of the experiments conducted on MRM.) presents total terms of input datasets, total matched terms with MRM and percentage of matched terms.

In order to evaluate the performance of best reference models on participating datasets, the experiments are conducted on five different batches of datasets as presented in (Appendix B). The two collaborating reference models are tested five different times for each variable. After that an average of scores for five round is taken and compared, the results of each round are presented separately. Figure Error! No text of specified style in document.1 illustrates the results of round one in which a comparison between the MRM and SRM for total terms and matched terms are discussed.

Fig. 4 depicts a significant result of the round 1 for all participating datasets using SRM and MRM. On left of the figure, the results of existing SRM and on the right the results of proposed MRM are shown. The first set of analysis begins with performance of SRM on participating datasets. Initially, 2564 terms of ACE2020 were tested on SRM, out of which 1212 were matched successfully. Secondly, 2192 terms of Aquaint dataset were examined, out of which only 551were matched. Similarly, 5740 terms of Sarcasm dataset were tested out of which merely 1198 were matched. Subsequently, 16 terms of UoA dataset were examined on SRM, out of which 15 were matched. Lastly, the 12 terms for HUA were tested and out of which 11 were matched. The results show a variation in matching of terms with the use of SRM, but it performed well on the UoA and HUA datasets.

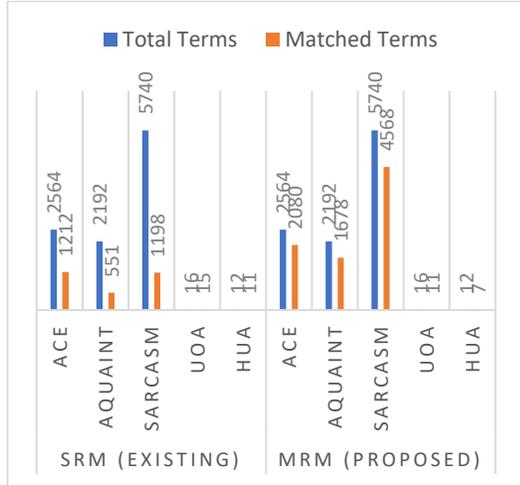

Fig. 4. Number of Succesfully Matched terms for MRM and SRm on batch 1

On the other hand of analysis, the input datasets are used to test the performance of MRM. At first, 2564 terms of ACE2020 were tested on MRM, out of which 2080 were matched magnificently. Subsequently, 2192 terms of Aquaint dataset were examined, out of which only 1678 were matched well. Similarly, 5740 terms of Sarcasm dataset were tested out of which 4568 were matched perfectly. Afterwards, 16 terms of UoA dataset were examined on MRM, out of which 11 were matched. Last of all, the 12 terms for HUA were tested and out of which 7 were matched.

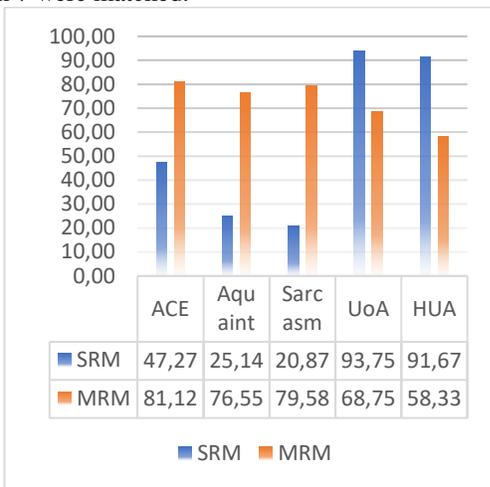

Fig. 5. Percentage of Matched terms for MRM and SRM on Batch 1

The performance findings from this round suggest that the MRM performed better than SRM on ACE 2020, Aquaint and Sarcasm datasets whereas the SRM works better on HUA and UoA datasets. Fig. 5 illustrates the terms matched (in terms of matched percentage) with both reference models on participating datasets.

The percentage of matched terms using SRM for ACE2020, Aquaint, Sarcasm, UoA and HUA are 47%, 25%, 20%, 93% and 91%, respectively. Whereas the percentage of matched terms using MRM for ACE2020, Aquaint, Sarcasm, UoA and HUA are 81%, 76%, 79%, 68% and 58%, respectively. This is because the proposed MRM covers multiple dimensions such as context, semantic and syntactic clues. One of the significant contributions of MRM is that it checks each participating word/token from input dataset with index based domain knowledge.

With that, the input tokens are checked multiple times and based on the context and similarity score of the index it produces very similar words. It is worth noting that if any of the tokens' score is high based on the similarity, but the score is less in terms of context than the terms which matched based on the context are selected. Whereas the existing SRM only checks the similarity based on string and lexical similarity and only in single dimension.

Comparative analysis on the results of SRM and MRM shows that the performance of MRM is better than the SRM on ACE 2020, Aquaint and Sarcasm datasets while the SRM performs better on HUA and UoA datasets. The results of MRM on UoA and HUA datasets are low which is due to different domain knowledge (medical) of the datasets. In Fig. 6, the performance of SRM and MRM are measured for batch 5 on contributing datasets. The remaining batches (2-4) are not presented here but the average of all five batches are presented in table 1. (Appendix B).

A significant result of the round 5 for all participating datasets using SRM and MRM. On left of the figure, the results of existing SRM and on the right the results of proposed MRM are shown. The first set of analysis begins with performance of SRM on participating datasets. Initially, 12820, terms of ACE2020 were tested on SRM, out of which 5605 were matched successfully. Secondly, 10960 terms of Aquaint dataset were examined, out of which only 2405 were matched. Similarly, 28700 terms of Sarcasm dataset were tested out of which merely 4701 were matched. Subsequently, 82 terms of UoA dataset were examined on SRM, out of which 70 were matched. Lastly, the 60 terms for HUA were tested and out of which 50 were matched. The results show a variations in matching of terms with the use of SRM, but it performed well on the UoA and HUA datasets.

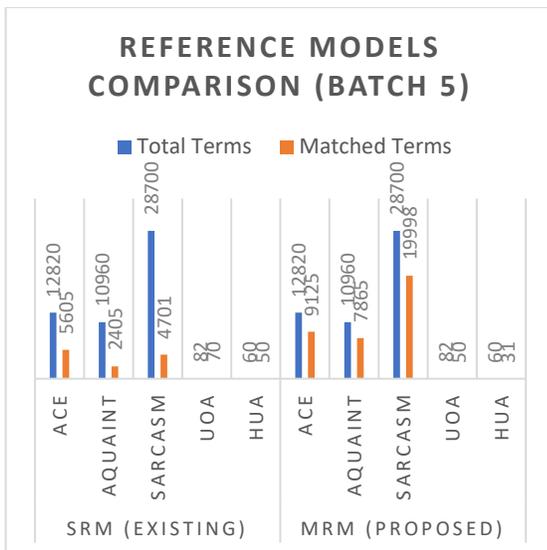

Fig. 6. Number of Succesfully Matched terms for MRM and SRM on Batch 5

On the other hand of analysis, the input datasets are used to test the performance of MRM. At first, 12820 terms of ACE2020 were tested on MRM, out of which 9125 were matched magnificently. Subsequently, 10960 terms of Aquaint dataset were examined, out of which only 7865 matched well. Similarly, 28700 terms of Sarcasm dataset were tested out of which 19998 were matched perfectly. Afterwards, 82 terms of UoA dataset were examined on SRM, out of which 50 were matched. Last of all, the 60 terms for HUA were tested and out of which 31 were matched.

For validation of performance, the SRM and MRM results are presented here. The comparison of performances (in terms of percentage) is depicted in Fig. 6.

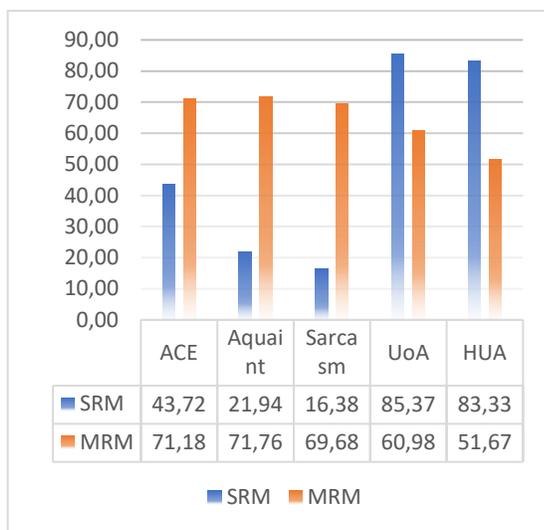

Fig. 7. Percentage of Matched terms for MRM and SRM on Batch 5

The performance findings from this round suggest that the MRM performed better than SRM on ACE 2020, Aquaint and Sarcasm datasets whereas the SRM works better on HUA and UoA datasets.

Fig. 7 illustrates the terms matched (in terms of matched percentage) with both reference models on participating datasets. The percentage of matched terms using SRM for ACE2020, Aquaint, Sarcasm, UoA and HUA are 44%, 22%, 17%, 85% and 83%, respectively. Whereas percentage of matched terms using MRM for ACE2020, Aquaint, Sarcasm, UoA and HUA are 71%, 72%, 70%, 61% and 52%, respectively. This is because the proposed MRM covers multiple dimensions such as context, semantic and syntactic clues.

One of the significant contributions of MRM is that it checks each participating word/token from input dataset in domain knowledge by adopting the functionality of indexing. With that, the input tokens are checked multiple times and based on the context and similarity score of the index. It's worth noting that that if any of the tokens' score is high based on the similarity but the score is less in terms of context then the terms which matched based on the context are selected. Whereas the existing SRM only checks the similarity based on string similarity and only in single dimension.

Comparative analysis on the results of SRM and MRM shows that the performance of MRM is better than the SRM on ACE 20202, Aquaint and Sarcasm datasets while the SRM performs better on HUA and UoA datasets. The results of MRM on UoA and HUA datasets are low which is due to different domain knowledge (medical) of the datasets.

## V. CONCLUSION AND FUTURE RECOMMENDATIONS

During the literature review and aiming to find solutions to solve the data heterogeneity, it was found that the only possible solution to solve the problem is to harmonize data. By adopting many techniques such as semantic, lexical matching and reference matching template. Based on that, a reference model which was developed by [14] for data curation framework for medical cohort taken as baseline study. In that, the reference model (SRM) contains the domain knowledge of specific terms that were used in medical domain. The performance of MRM has been evaluated on five heterogeneous (structured, semi-structured and unstructured) datasets and in five multiple rounds. The results of each rounds of ACE20220, Aquaint, Sarcasm, UoA and HUA show better performance of MRM over its counterpart reference model i.e., Single dimensional reference model. The overall performance of MRM on all participating datasets is more than 30% on ACE2020, Aquaint, and Sarcasm datasets whereas the performance of UoA and HUA performed better on SRM. To conclude with the performance of MRM, it has been observed that the use of MRM supports the DHF in selection of key terms based on semantic and lexical matched terms. *Contribution:* Design and development of Multidimensional Reference Model which is developed based on the linguistics categories such as context, semantic and syntactic clues. The model enables the use of indexing for any English sentences by introducing the words and their respective score. The proposed MRM produced huge number of words that can be used as a reference for any general domain which contains daily basis data generated in textual formats. *Future Recommendation:* Use of other categories of linguistics and computational linguistics for further improvement in the field of English grammar.

ACKNOWLEDGMENT

This paper was fully supported by Universiti Teknologi PETRONAS.

Appendix A

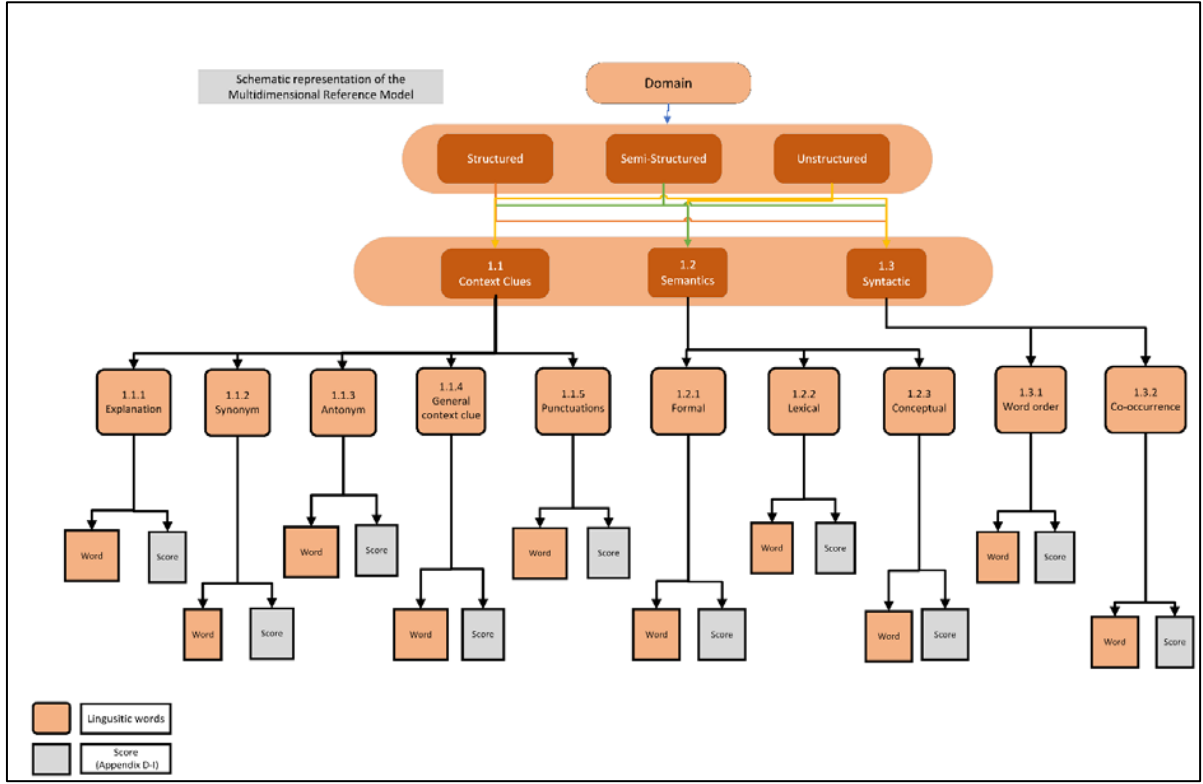

Fig. 8. Percentage of Matched terms for MRM and SRM on Batch 5

Appendix B

TABLE I. TABLE TYPE STYLES

| Datasets | Reference Models  Performances Batches | SRM | | | MRM | | |
|---|---|---|---|---|---|---|---|
| | | Total terms | Total Matched terms | Matched terms % | Total terms | Total Matched terms | Matched terms % |
| ACE 2020 | Batch 1 | 2564 | 1212 | 47.27 | 2564 | 2080 | 81.12 |
| | Batch 5 | 12820 | 5605 | 43.72 | 12820 | 9125 | 71.18 |
| | **Average** | **7692** | **3457.60** | **45.57** | **7692** | **5689.4** | **75.64** |
| Aquaint | Batch 1 | 2192 | 551 | 25.14 | 2192 | 1678 | 76.55 |
| | Batch 5 | 10960 | 2405 | 21.94 | 10960 | 7865 | 71.76 |
| | **Average** | **6576** | **1511.20** | **23.57** | **6576** | **4797.4** | **73.68** |
| Sarcasm | Batch 1 | 5740 | 1198 | 20.87 | 5740 | 4568 | 79.58 |
| | Batch 5 | 28700 | 4701 | 16.38 | 28700 | 19998 | 69.68 |
| | **Average** | **17320** | **3108.40** | **18.69** | **17320** | **12686.8** | **75.01** |
| UoA | Batch 1 | 16 | 15 | 93.75 | 16 | 11 | 68.75 |
| | Batch 5 | 82 | 70 | 85.37 | 82 | 50 | 60.98 |
| | **Average** | **49** | **43.00** | **88.83** | **49** | **30.6** | **63.34** |
| HUA | Batch 1 | 12 | 11 | 91.67 | 12 | 7 | 58.33 |
| | Batch 5 | 60 | 50 | 83.33 | 60 | 31 | 51.67 |
| | **Average** | **36** | **31.40** | **88.61** | **36** | **19.2** | **54.22** |

Appendix C,D

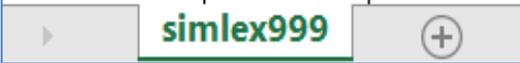

Samples of Linguistics Context (Synonyms and Antonyms)

Appendix E,F

| | | | |
|---|---|---|---|
| | Obama to name Susan Rice as national security adviser | Donilon out, Rice in as Obama's national security adviser | |
| 1.3 | The foundations of South Africa are built on Nelson Mandela's memory | Australian politicians lament over Nelson Mandela's death | |
| 3 | Turkish riot police tear gas Taksim Square protest | Turkish riot police enter Taksim Square | |
| 5 | Chicago Shooting Shows Gap in Stepped-up Policing | Chicago shooting shows gap in stepped-up policing | |
| 1.2 | Ukraine protest leaders name ministers, Russian troops on alert | Ukraine Refuses to Act Against Russian 'Provocation' | |
| 1.2 | North Korea shuns offer of talks | North Korea shoots 2 rockets | |
| 5 | Prince Charles 'compares Putin to Hitler' | Prince Charles 'compares Vladimir Putin to Adolf Hitler' | |
| 3.6 | Malala, Snowden, Belarusians Short-Listed For Sakharov Prize | Fugitive Snowden short-listed for European rights prize | |
| 1.2 | Israel PM accuses Iran president of hypocrisy | Greek PM accuses coalition of hypocrisy | |
| 3.2 | One dead in Philadelphia building collapse, others pulled from rubble | Six confirmed dead after Philadelphia building collapse | Samples of Lexical Semantics (Formal) |
| 2.8 | Police clash with youth in Cairo after anti-Morsi protest | Egypt: Police clash with pro-Morsi protesters | |
| 2.4 | Top Diplomats Meet in Munich at Critical Time | Top Diplomats Discuss Iran's Nuclear Program | |
| 1.2 | Snowden 'given refugee document by Ecuador' | Snowden poised to fly out of Moscow | |
| 2.2 | NA President,Äôs message on the World Press Freedom Day | Pakistan marks World Press Freedom Day | |
| 3.6 | EU foreign ministers to discuss Syria arms embargo | EU foreign ministers seek solution on Syria arms | |
| 3.6 | Death toll from Philippine earthquake rises to 185 | Death toll from Philippines quake rises to 144 | |
| 4 | Stocks to watch at close on Monday | Stocks to watch on Monday | |
| 3.2 | Army jets kill 38 militants in NW Pakistan air raids | U.S. drone kills 4 militants in Pakistan | |
| 4 | Egyptian police fire tear gas at protesters in Cairo | Police fire tear gas at protesters in Cairo | |
| 3.2 | ElBaradei to Become Egyptian PM | Liberal ElBaradei named Egypt PM, Islamists cry foul | |
| 1.4 | Philippines holds second senator over corruption | Philippines recovering after powerful typhoon | |
| 1.4 | German envoy optimistic about Iran-G5+1 talks | Iran 'cautiously optimistic' about future nuclear talks | |
| 4.6 | Renowned Spanish flamenco guitarist Paco De Lucia dies | Spanish flamenco guitarist Paco de Lucia dies at 66 | |
| 2 | Chinese icebreaker changes course towards suspicious objects | Chinese search plane finds 'suspicious objects' | |
| 5 | Snowden Hits Hurdles in Search for Asylum | Snowden's hits hurdles in search for asylum | |
| 2.8 | Death toll in building collapse in south India mounts to 47 | 4 killed in building collapse in southern India | |
| 1.4 | Pakistan imposes temporary ban on 2 TV channels | Pakistan Dismisses Case against FBI Agent | |
| 1 | Palestinian prisoners arrive at Muqata in Ramallah | Pakistani prisoner assaulted in Jammu jail | |
| 1 | 10 dead, five injured in SW China road accident | 5 hurt in Gaza City car accident | |
| 5 | Matt Smith quits BBC,Äôs Doctor Who | Matt Smith quits BBC's Doctor Who | |
| 1.8 | Thai junta amasses security force to smother Bangkok protests | Thai junta security forces stay in barracks as protests dwindle | |
| 1 | Queen pays tribute to Nelson Mandela | South Africa's rugby fraternity mourns Mandela | |
| 0.4 | Declines in US stock market moderate | Tech sell-off sends Asian stock markets lower | |
| 2.6 | Latest Anti-Muslim Violence in Burma Kills 1, Injures 10 | World Briefing \| Asia: Myanmar: Deadly Anti-Muslim Violence Flares Up | |
| 3.8 | Suspected U.S. drone strike kills 5 in Pakistan | U.S. drone strike kills 5 in Pakistan | |
| 2 | The Note's Must-Reads for Friday May 24, 2013 | The Note's Must-Reads for Tuesday October 29, 2013 | |
| 0.6 | 22 killed in mine accident in southwestern China | 2 killed, dozens injured by blast in southwest Pakistan | |
| 4.8 | NYPD Twitter Outreach Backfires Badly | NYPD's twitter campaign backfires | |
| 4.4 | 19 hurt in New Orleans shooting | Police: 19 hurt in NOLA Mother's Day shooting | |
| 3.8 | Another migrant ship capsizes off Italy | Another migrant boat capsizes off Italy, 27 dead | |
| 5 | Turkish search ends as last missing miners found | Turkish Search Ends as Last Missing Miners Found | |
| 1.4 | Iran predicts failure of Israeli-Palestinian peace talks | Tentative Deal Reached to Resume Israeli-Palestinian Talks - US | |
| 4.4 | Olivia Colman wins second BAFTA Award | Baftas 2013: Olivia Colman picks up two awards | Samples of Lexical Semantics (Lexical) |
| 3.8 | Titanic Violin Nabs Record $1.4 Million | Titanic violin sells for $1.7 million | |
| 5 | Ankeet Chavan granted conditional bail for marriage | Ankeet Chavan granted bail to get married | |
| 3 | JAL's first order from Airbus is blow to Boeing | Japan Airlines orders 31 Airbus A350s valued at $9.5 bn | |
| 3 | Sienna Miller testifies at UK phone hacking trial | Sienna Miller attacks press for ,Äëtitillating,Äô reports at hacking trial | |
| 2.6 | Israelis attack 2 Palestinians in Jerusalem area | Settlers Beat up Palestinian in Jerusalem | |
| 0.4 | EU extends sanctions against Russia | Nigeria drops arms trafficking charges against Russian sailors | |
| 3.4 | ,ÄêGlee,Äô star Cory Monteith found dead in hotel room | Cory Monteith found dead: Canadian ,ÄêGlee,Äô star was 31 | |
| 4.6 | Lebanon's PM forms 'unity cabinet' Lebanon's prime minister has formed a cabinet more than 10 months after taking office, taking in a wide range of political groups after bridging serious divisions am | BEIRUT: Lebanon's prime minister formed a cabinet more than 10 months after taking office yesterday, including a wide range of political groups after bridging serious divisions among them mostly over | |
| 0.6 | Iranian president makes debut on world stage | broken | |
| 1.2 | Scores Killed In Egyptian Protests | Turkey's PM Warns Against Protests | |
| 3.8 | State Dept. issues wide travel alert, says terror attack possible | US issues global travel alert, cites al-Qaida threat | |
| 4.6 | Michelle Obama To Star In Parks And Recreation | Michelle Obama to appear on 'Parks and Recreation' | |
| 1.6 | Singapore stocks end up 0.11 pct | Singapore stocks end down 0.45% | |
| 4.6 | World's oldest man dies at 116 | World's oldest ever man dies aged 116 | |
| 2.8 | Death toll in Lebanon bombings rises to 47 | 1 suspect arrested after Lebanon car bombings kill 45 | |
| 5 | Greek far-right leader imprisoned pending trial | Greek Far-Right Leader Imprisoned Pending Trial | |

Appendix G

| | | | |
|---|---|---|---|
| you know | INT | 27348 | The big step is then getting the rest of the Council to take it on board, that's the big step, …the development budget there went on projects for young people, you know, so there |
| I think (that) | CA | 25862 | I think I've still got the piece about that.. |
| a bit | NP | 7766 | Can you move round this side a bit? |
| (always [155], never [87]) used to {INF} | VP | 7663 | We used to look forward to them coming. |
| as well | AVP | 5754 | Can I just say something else as well? |
| a lot of {N} | NP | 5750 | I got a lot of letters from the children there and which was very gratifying. |
| {No.}pounds | NP | 5598 | I'd also ask you to consider costs of ten pounds. |
| thank you | VP | 4789 | Well thank you for that that's a very good start to the evening |
| {No.}years | NP | 4237 | I've done it for seven years. |
| in fact | PP | 3009 | In fact, if the previous speaker has complained about waiting in patience, I have waited forty years to tell this story in the assembly… |
| very much | AVP | 2818 | …they enjoy it very much… |
| {No.}pound | NP | 2719 | I've got to take a taxi of one pound forty a day to shop… |
| talking about {sth} | VP | 2489 | It was a different from what you're talking about. |
| (about [91.]) {No.} percent {of sth [580]}, in sth [ | NP | 2312 | As I said, we've already got forty one percent of themain funding required for the project… |
| I suppose (that) | VP | 2281 | I suppose that was one way of nothing being done. |
| at the moment | PP | 2176 | Well I haven't done anything at the moment because I didn't think it was worth it actually. |
| a little bit | NP | 1935 | And the other times your concentration will drop a little bit |
| looking at {sth} | VP | 1849 | I think there's another way of looking at that. |
| this morning | AVP/NP | 1846 | Oh she was screaming this morning. |
| (not) any more | AP | 1793 | …women shouldn't have off days any more. |
| come on | INT | 1778 | No, it's not a verb, come on, what is it? |
| number{No.} | NP | 1661 | Number six which is the thing that we have to look at. |
| come in (swc, sth) | VP | 1571 | We're about to finish, so please come in. |
| come back | VP | 1547 | We'll come back to you in a second. |
| have a look | VP | 1471 | You can go and have a look. |
| in terms of {sth} | PP | 1403 | I think it was one of the things which never really took off in terms of the accident. |
| last year | AVP/NP | 1347 | That was last year or was that two years ago? |
| so much | AP/AVP | 1334 | He loved the sea so much. |
| {No.}years ago | AVP | 1314 | That was last year or was that two years ago? |

Samples of word-order

Appendix H

| | | | | |
|---|---|---|---|---|
| 1. well | as well | AVP | 5754 | Can I just say something else as well? |
| | very well | AP/AVP | 987 | You can read very well can't you? |
| | as well as | C/P | 620 | …other schools in the area used to use this facility as well as we did. |
| | well done | VP | 171 | Top of the class, well done. |
| | really well | AVP | 118 | I've done really well. |
| | quite well | AP | 94 | I thought it was quite well organized considering. |
| | so well | AP | 80 | …got the final ball wrong but a shame he'd done so well. |
| | well known | VP | 74 | I mean it's more well known than it used to be. |
| | very well | INT | 68 | Very well, thanks |
| | here as well | AVP | 47 | Feels warm in here as well. |
| | pretty well | AVP | 43 | Dave's got a job pretty well hasn't he? |
| | very well | AP | 12 | he's not very well, he looks how pale he is, James is quite pale as well |
| 2. know | you know | INT | 27348 | The big step is then getting the rest of the Council to take it on board, that's the big step, …the development budget there went on projects for young peo |
| | know that {S V} | VP | 889 | …you know that this is the only room available. |
| | know it | VP | 469 | It doesn't really worry me whether you know it or not. |
| | know if {S V} | VP | 321 | I don't know if any of you are old enough to remember… |
| | know whether {S V} | VP | 247 | I don't know whether anybody would disagree with that. |
| | as I know | CA | 95 | …just as I know I must call you my lady Anne! |
| | know this | VP | 88 | …issue of drawings to the client wished to know, wished to know this. |
| | (not) know anything | VP | 86 | …I don't really know anything special about me. |
| | know the one | VP | 56 | Ruth, you know the one I used to look after? |
| | know the answer | VP | 47 | Well I supposed we'd all like to know the answer to that. |
| | know one | VP | 42 | I know one who wouldn't stand for it. |

Samples of co-occurrence words